\documentclass{mymanuscript} 

\usepackage{amsmath,amsfonts,amssymb}
\usepackage[caption=false,font=normalsize,labelfont=sf,textfont=sf]{subfig}
\usepackage{siunitx}
\usepackage{slashed}

\begin{document}

\articletype{Paper}

\title{Dynamic Doppler Effects on Dirac Spinor Fields}

\author{Bryce M. Barclay$^{1,*}$ and Alex Mahalov$^{1}$}

\affil{$^1$School of Mathematical and Statistical Sciences, Arizona State University, Tempe, USA}

\affil{$^*$Author to whom any correspondence should be addressed.}

\email{Bryce.Barclay@asu.edu}


\begin{abstract}
    We derive the dynamic Doppler effects of noninertial observer motion on Dirac spinor fields in 4D spacetime, characterizing how proper acceleration, proper jolt (also known as jerk), observer path curvature, torsion, and hyper-torsion modify the inertial quantum description of free particles. 
    We delineate the impacts of nonlinear observer motion on the amplitude and phase of the Dirac spinor wave field and isolate the spin-induced signatures that have no counterpart in the scalar Klein-Gordon field.
    For worldlines with constant proper jolt, the wave-field amplitude scales with quadratic rapidity. 
    The complex Frenet-Serret curvature invariants generate both an exponential amplitude factor and a spin-induced phase factor. 
    Spinor structure therefore produces an observable dynamic Doppler signature distinct from any scalar prediction. 
    We analyze the effects of proper jolt and nonlinear rapidity on quantized Dirac fields in the context of acceleration-induced $\beta^+$-decay and derive the instantaneous decay rate, an equivalent of decay rate for non-stationary motions. 
    This work provides new insight for noninertial-proton $\beta^+$-decay and a general framework for testable variable-rapidity Unruh predictions. 
\end{abstract}

\section{Introduction}

The effects of noninertial motion on observation and detection in quantum systems have important implications for quantum physics and provide a wealth of potential applications.
The Unruh effect \cite{Unruh1976,dickinson2025new}, which describes the apparent change in quantum fields due to noninertial motion, has application in quantum information and quantum computing \cite{lemaitre2025universal}, condensed matter physics \cite{iorio2014quantum}, and quantum optics \cite{chen1999testing,schutzhold2006signatures}. 
Many modern commercial applications require acceleration of quantum particles to relativistic velocities in particle beams \cite{sheehy2024applications}. 
Analysis of the quantum effects of noninertial motion is also important for the development of particle detection frameworks, such as the Unruh-DeWitt detector 
\cite{HummerMartinMartinezKempf2016,LoukoToussaint2016,sachs2017entanglement,letaw1981stationary}.
The framework of quantum (Unruh-DeWitt) detectors and accelerating observers has also been studied in the context of spinor fields \cite{WuTangShi2023,alsing2006entanglement,Mashhoon2007}. 
The Unruh effect~\cite{Unruh1976,dickinson2025new}, in which a uniformly accelerated observer perceives the Minkowski vacuum as a thermal state, illustrates that the particle content of a quantum field is sensitive to the worldline along which the field is interrogated. A complementary illustration has recently been given in \cite{rukan2026truncated}, whose analysis of a photon truncated by an optical shutter shows that even a locally defined operation can yield a state that is globally a superposition over photon number from zero to infinity while remaining locally equivalent to a single photon and to the vacuum. 
Consequences of the Unruh effect, such as the decay of protons due to acceleration \cite{CrispinoHiguchiMatsas2008}, have the potential to alter our understanding of subatomic particles.

The role of acceleration as a control parameter for complex dynamics is a recurring theme across classical hydrodynamics. 
In \cite{narasimha1973relaminarization}, the authors showed how acceleration can impact turbulence, with pressure forces coming to dominate slowly responding Reynolds stresses.
These considerations have recently been extended to Rayleigh-Taylor and Richtmyer-Meshkov mixing driven by accelerations \cite{abarzhi2022self,abarzhi2026scale}.  
The present paper studies the analogous role of constant and time-varying acceleration on a relativistic quantum field, where the central kinematic variables are the Frenet-Serret curvatures of the observer's worldline and the proper jolt \cite{russo2009relativistic}.

Here we investigate the core effects of moving references frames on quantum dynamics and derive new results for higher-order kinematic moments and geometric curvature parameters. 
Our theory includes nonlinear trajectories with acceleration that changes in both direction and magnitude. 
We incorporate the effects of spin by analyzing Dirac spinor fields as opposed to the scalar Klein-Gordon field. 
A stochastic counterpart to the amplitude-phase perturbations derived here arises in semiconductor laser sources, where spontaneous-emission and carrier noise admit a unified stochastic differential equation description \cite{mcdaniel2018stochastic}.

Concretely, this paper makes five contributions.
First, we extend the noninertial Dirac wave function from the Rindler case to an arbitrary stationary worldline parameterized by Frenet-Serret curvature, torsion, and hyper-torsion.
Second, we treat time-dependent rapidity, including a closed-form result for constant proper jolt (also known as jerk).
Third, we identify a spin-induced phase contribution $\tau\omega/2$ and a spin-state-selective amplitude response, both absent in the Klein-Gordon case.
Fourth, we cast the analogy with classical dynamic Doppler effects into a quantitative form that can be compared with laser-driven, condensed-matter, and Unruh-DeWitt-detector experiments on fermionic systems.
Fifth, we show how the prediction of the weak decay of accelerating protons can be generalized to protons with nonlinear rapidity in the context of quantum field theory. 

\section{The Dirac equation and spinor representations of Lorentz transformations}

The Dirac equation, 
\begin{align}\label{Dirac_equation}
    (i\gamma^\mu\partial_\mu - m)\psi(x) = 0,
\end{align}
is a first-order PDE in Minkowski spacetime whose solution $\psi(x)$, called the Dirac spinor field, is a complex 4-vector at every point. 
The parameter $m$ is the mass of a spin-$1/2$ particle and $\gamma$ is a Dirac algebra 4-vector which does not depend on the location $x$ in spacetime and satisfies the relation 
\begin{align}
    \{\gamma^\mu,\gamma^\nu\} := \gamma^\mu\gamma^\nu + \gamma^\nu\gamma^\mu = 2\eta^{\mu\nu}I_4. 
\end{align}
The Weyl representation of the gamma matrices is  
\begin{align}
    \gamma^0 = 
    \begin{bmatrix}
        0 & I \\
        I & 0
    \end{bmatrix},\qquad 
    \gamma^i = 
    \begin{bmatrix}
        0 & \sigma_i \\
        -\sigma_i & 0
    \end{bmatrix}.
\end{align}

The Dirac adjoint $\overline{\psi} = \psi^\dagger\gamma^0$ satisfies the adjoint equation 
\begin{align}
    \overline{\psi}(x)(-i\gamma^\mu\partial_\mu - m) = 0,
\end{align}
where, in this case, $\partial_\mu$ are understood as the partial derivative operators acting from the right. 
The Dirac adjoint gives us the Lorentz invariant quantity $\overline{\psi}\psi$ and is necessary to define the Lagrangian density for the Dirac equation 
\begin{align}
    \mathcal{L} = \overline{\psi}(i\gamma^\mu\partial_\mu - m)\psi.
\end{align}
The Dirac field has the conserved current
\begin{align}
    J^\mu &= \overline{\psi}\gamma^\mu \psi 
\end{align}
which arises from the continuous symmetry of the wave function phase.

Lastly, note that the Klein-Gordon equation 
\begin{align}\label{Klein_Gordon_equation}
    (\partial^2 + m^2)\psi = 0,
\end{align}
can be derived from the Dirac equation (\ref{Dirac_equation}): 
\begin{align}
    (-i\gamma^\mu\partial_\mu - m)(i\gamma^\nu\partial_\nu - m)\psi 
    = (\frac{1}{2}\{\gamma^\mu,\gamma^\nu\}\partial_\mu\partial_\nu + m^2)\psi 
    = (\partial^2 + m^2)\psi. \nonumber
\end{align}

The Dirac field is a spinor field, which under a Lorentz transformation $x^\mu \to {x'}^\mu = {\Lambda^\mu}_\nu x^\nu$ transforms as
\begin{align}
    \psi(x) \to \psi'(x') = D[\Lambda]\psi(\Lambda^{-1}x'),
\end{align}
where $D[\Lambda]$ is the spinor representation of $\Lambda$.
Writing $\Lambda$ and $D[\Lambda]$ as exponentials of antisymmetric parameters $\Omega_{\rho\sigma}$,
\begin{align}
    \Lambda = \exp\left(-\tfrac{i}{2}\Omega_{\rho\sigma}\mathcal{M}^{\rho\sigma}\right),
    \qquad
    D[\Lambda] = \exp\left(-\tfrac{i}{2}\Omega_{\rho\sigma}S^{\rho\sigma}\right),
\end{align}
the generators are
\begin{align}
    \left(\mathcal{M}^{\rho\sigma}\right)_{\mu\nu} = i({\delta^\rho}_\mu{\delta^\sigma}_\nu - {\delta^\rho}_\nu{\delta^\sigma}_\mu),
    \qquad
    S^{\rho\sigma} = \frac{i}{4}[\gamma^\rho,\gamma^\sigma].
\end{align}
In the Weyl basis the spinor generators take block-diagonal form,
\begin{align}\label{eq:S_blocks}
    S^{0i} = -\frac{i}{2}
    \begin{bmatrix}
        \sigma_i & 0 \\
        0 & -\sigma_i
    \end{bmatrix}, \qquad
    S^{ij} = \frac{\epsilon^{ijk}}{2}
    \begin{bmatrix}
        \sigma_k & 0 \\
        0 & \sigma_k
    \end{bmatrix},
\end{align}
so that a pure boost with rapidity 3-vector $\chi$ and a pure rotation with angle 3-vector $\theta$ act on the upper-Weyl two-spinor as
\begin{align}\label{eq:boost_rotation_2spinors}
    D[\Lambda_B] =
    \begin{bmatrix}
        \exp(\chi\cdot\sigma/2) & 0 \\
        0 & \exp(-\chi\cdot\sigma/2)
    \end{bmatrix},
    \quad
    D[\Lambda_R] =
    \begin{bmatrix}
        \exp(i\theta\cdot\sigma/2) & 0 \\
        0 & \exp(i\theta\cdot\sigma/2)
    \end{bmatrix},
\end{align}
where $\sigma = (\sigma_1,\sigma_2,\sigma_3)$.
Using the identity $(\hat{n}\cdot\sigma)^{2} = I_2$ for any unit 3-vector $\hat n$, the exponentials reduce to elementary functions: $\exp(i\theta_0\hat{n}\cdot\sigma/2) = I_2\cos(\theta_0/2) + i\hat{n}\cdot\sigma\sin(\theta_0/2)$ and $\exp(\chi_0\hat{n}\cdot\sigma/2) = I_2\cosh(\chi_0/2) + \hat{n}\cdot\sigma\sinh(\chi_0/2)$.

More generally, the matrix $A = a_1\sigma_1 + a_2\sigma_2 + a_3\sigma_3$ with $a_i\in\mathbb{C}$ satisfies $A^2 = (a_1^2+a_2^2+a_3^2)I_2$ by the Cayley-Hamilton theorem, from which
\begin{align}\label{eq:cayley_hamilton_exp}
    \exp(A) = I_2\cosh\!\sqrt{a_1^2+a_2^2+a_3^2} + A\,\frac{\sinh\!\sqrt{a_1^2+a_2^2+a_3^2}}{\sqrt{a_1^2+a_2^2+a_3^2}}.
\end{align}
Equation~(\ref{eq:cayley_hamilton_exp}) is the key identity behind our closed-form Frenet-Serret results in Section~\ref{sec:frenet} since it remains valid for complex $a_i$ and so handles mixed boost-rotation generators in a single expression.

\section{Dirac spinor fields in variable-acceleration reference frames: observer jolt and path curvatures}

To obtain the Dirac spinor in a non-inertial frame of reference, we use the \textit{locality hypothesis} which says that the reference frame of a non-inertial observer is given by the instantaneous co-moving frame. 

\subsection{Variable acceleration and higher-order kinematics}

We restrict attention here to motion along the $x^1$ direction with rapidity
\begin{align}\label{eq:chi_def}
    \chi(\tau) = \chi_0 + a_0\tau + j_0\tau^2/2,
\end{align}
where $\tanh(\chi_0) = \beta = v_0/c$ is fixed by the initial velocity $v_0$, $a_0 = |a(0)|$ is the magnitude of the initial proper acceleration, and $j_0 = |\sigma|$ is the magnitude of the (constant) proper jolt $\sigma^\mu = d a^\mu/d\tau - a^\nu a_\nu u^\mu$ \cite{russo2009relativistic}. The quantities $u$ and $a$ denote the 4-velocity and 4-acceleration of the observer, respectively. 
From an inertial frame with coordinates $(x^\mu)$, the worldline of the accelerating observer is then 
\begin{align}
    z^0(\tau) &= \int_0^{\tau} \cosh(\chi(\tau'))\,d\tau',\label{rela_jolt_path0}\\
    z^1(\tau) &= \int_0^{\tau} \sinh(\chi(\tau'))\,d\tau' + z^1(0),\label{rela_jolt_path1}
\end{align}
which collects both the initial velocity and the time-dependent rapidity into the single function $\chi(\tau)$.
The instantaneous co-moving frame is obtained from this worldline; the corresponding spinor transformation follows from it.

The instantaneous co-moving frame is obtained via the Lorentz boost
\begin{align}
    \Lambda(\tau) = 
    \begin{bmatrix}
        \cosh(\chi(\tau)) & \sinh(\chi(\tau)) & 0 & 0 \\
        \sinh(\chi(\tau)) & \cosh(\chi(\tau)) & 0 & 0 \\
        0 & 0 & 1 & 0 \\
        0 & 0 & 0 & 1 
    \end{bmatrix}
    =
    \exp\left(-\frac{i}{2}\left(\chi(\tau)\mathcal{M}^{01} + (-\chi(\tau))\mathcal{M}^{10}\right)\right). 
\end{align}
Therefore, the spinor transformation is given by 
\begin{align}
    D[\Lambda(\tau)] 
    &= 
    \exp\left(-\frac{i}{2}\left(\chi(\tau)S^{01} + (-\chi(\tau))S^{10}\right)\right)\\
    &=
    \begin{bmatrix}
        \cosh(\chi(\tau)/2) & -\sinh(\chi(\tau)/2) & 0 & 0 \\
        -\sinh(\chi(\tau)/2) & \cosh(\chi(\tau)/2) & 0 & 0 \\
        0 & 0 & \cosh(\chi(\tau)/2) & \sinh(\chi(\tau)/2) \\
        0 & 0 & \sinh(\chi(\tau)/2) & \cosh(\chi(\tau)/2)
    \end{bmatrix}.
\end{align}

The eigenvalues of this matrix are 
\begin{align}
    \lambda_\pm = \cosh(\chi(\tau)/2) \pm \sinh(\chi(\tau)/2) = \exp(\pm\chi(\tau)/2).
\end{align}
The eigenvectors are 
\begin{align}
    v_{-,1} &= \frac{1}{\sqrt{2}}
    \begin{bmatrix}
        1 & 1 & 0 & 0
    \end{bmatrix}^T\\
    v_{+,1} &= \frac{1}{\sqrt{2}}
    \begin{bmatrix}
        -1 & 1 & 0 & 0
    \end{bmatrix}^T\\
    v_{-,2} &= \frac{1}{\sqrt{2}}
    \begin{bmatrix}
        0 & 0 & 1 & -1
    \end{bmatrix}^T\\
    v_{+,2} &= \frac{1}{\sqrt{2}}
    \begin{bmatrix}
        0 & 0 & 1 & 1
    \end{bmatrix}^T.
\end{align}
The components of the Dirac spinor are in terms of $z$ spin; however, the eigenvectors can be written more simply in terms of the $x$ spin basis: 
\begin{align}
    V &=
    \begin{bmatrix}
        \sigma_z & 0 \\
        0 & \sigma_x
    \end{bmatrix}
\end{align}
where $\sigma_x$ and $\sigma_z$ are the Pauli matrices in standard form. 

\subsection{The Frenet-Serret frame for nonlinear trajectories}\label{sec:frenet}

We compute the transformation of the spinor field from an inertial observer to an observer moving along a path with constant curvature, torsion, and hyper-torsion.

The Frenet-Serret frame $(e_\mu)$ in 4D Minkowski spacetime is defined by the equations 
\begin{align}
    \frac{de_0}{d\tau} &= \kappa_1e_1\\
    \frac{de_1}{d\tau} &= \kappa_1e_0 + \kappa_2 e_2\\
    \frac{de_2}{d\tau} &= -\kappa_2e_1 + \kappa_3 e_3\\
    \frac{de_3}{d\tau} &= -\kappa_3e_2,
\end{align} 
with $e_0$ corresponding to the 4-velocity of the observer. 
Written compactly:
\begin{align}
    \frac{de_\mu}{d\tau} = {(A^T)_\mu}^\nu e_\nu.
\end{align}
The Lorentz transformations are 
\begin{align}
    {\Lambda(\tau)^\mu}_\nu = \exp(\tau {A^\mu}_\nu) = \exp(\tau {(A^T)_\nu}^\mu) = \exp\left(-\frac{i}{2}\Omega_{\rho\sigma}\mathcal{M}^{\rho\sigma}\right).
\end{align}
Therefore $\Omega_{01} = \kappa_1 \tau$, $\Omega_{12} = -\kappa_2 \tau$, and $\Omega_{23} = -\kappa_3 \tau$, and the upper $2\times 2$ block of $D[\Lambda] = \exp(-\tfrac{i}{2}\Omega_{\rho\sigma}S^{\rho\sigma})$ takes the form $\exp(\tau A/2)$ with the complex Pauli-matrix combination
\begin{align}\label{eq:complex_A}
    A = -[(\kappa_1 - i\kappa_3)\sigma_1 - i\kappa_2\sigma_3] = \tilde{\kappa}\cdot\sigma,
\end{align}
where we have defined the complex 3-vector
\begin{align}\label{eq:kappa_tilde}
    \tilde{\kappa} = (-\kappa_1 + i\kappa_3,\,0,\,i\kappa_2).
\end{align}
Equation~(\ref{eq:cayley_hamilton_exp}) applies directly with $a_i = \tilde{\kappa}_i$, so
\begin{align}\label{eq:kappa_hat}
    \exp\!\left(\tfrac{\tau}{2}\tilde{\kappa}\cdot\sigma\right)
    = I_2\cosh\!\left(\tfrac{\tau}{2}\hat{\kappa}\right) + \frac{\tilde{\kappa}\cdot\sigma}{\hat{\kappa}}\sinh\!\left(\tfrac{\tau}{2}\hat{\kappa}\right),
    \qquad
    \hat{\kappa} = \sqrt{\tilde{\kappa}\cdot\tilde{\kappa}} = \sqrt{(-\kappa_1 + i\kappa_3)^2 + (i\kappa_2)^2}.
\end{align}
Writing out the matrix components with $\tilde{\kappa}_2 = 0$ gives the explicit upper block
\begin{align}\label{eq:D_upper_block}
    D[\Lambda]_{\rm upper}
    =
    \begin{bmatrix}
        \cosh\!\left(\tfrac{\tau}{2}\hat{\kappa}\right) + \tfrac{\tilde{\kappa}_3}{\hat{\kappa}}\sinh\!\left(\tfrac{\tau}{2}\hat{\kappa}\right) & \tfrac{\tilde{\kappa}_1}{\hat{\kappa}}\sinh\!\left(\tfrac{\tau}{2}\hat{\kappa}\right) \\
        \tfrac{\tilde{\kappa}_1}{\hat{\kappa}}\sinh\!\left(\tfrac{\tau}{2}\hat{\kappa}\right) & \cosh\!\left(\tfrac{\tau}{2}\hat{\kappa}\right) - \tfrac{\tilde{\kappa}_3}{\hat{\kappa}}\sinh\!\left(\tfrac{\tau}{2}\hat{\kappa}\right)
    \end{bmatrix}.
\end{align}
The lower $2\times 2$ block of $D[\Lambda]$ has the same form with the sign of the boost-generating terms reversed, following the block structure of (\ref{eq:S_blocks}).
The eigenvalues of the $2\times 2$ block of $D[\Lambda]$ are 
\begin{align}
    \lambda_\pm &= 
    \exp \left(\pm\frac{1}{2} \hat{\kappa} \tau \right).
\end{align}

To obtain the coordinates $(x'^\mu)$ of the noninertial observer, we first use the Frenet-Serret frame $e_\mu(\tau) = {\Lambda(\tau)_\mu}^\nu$ of the observer and integrate the velocity $e_0$ to get the world line of the observer:
\begin{align}
    z(\tau) - z(0) = \int_0^{\tau} e_0(\tau') \, d\tau'.
\end{align}
The coordinates of the observer are $(\tau, x'^1, x'^2, x'^3)$ which are related to the inertial coordinates via 
\begin{align}
    x^\mu = x'^1e_1^\mu(\tau) + x'^2e_2^\mu(\tau) + x'^3e_3^\mu(\tau) + z^\mu(\tau). 
\end{align}
We therefore compute ${\Lambda(\tau)_\mu}^\nu = \exp(\tau {A_\mu}^\nu)$. 
To calculate the exact exponential, we use an eigenvalue decomposition of $A$:
\begin{align}
    \exp(\tau A)
    =
    V\exp(\tau D_A) \eta V^T\eta.
\end{align}
The eigenvalues of $A$ are 
\begin{align}
    \lambda_{\pm,\pm} = \pm \sqrt{\lambda_\pm}, \qquad \lambda_\pm = L_1 \pm \sqrt{L_1^2 + L_2^2} \qquad L_1 = \frac{1}{2}(\kappa_1^2 - \kappa_2^2 - \kappa_3^2) \quad L_2 = \kappa_1\kappa_3
\end{align}
which can also be written $\pm \sqrt{\chi^2},\pm i\sqrt{\omega^2}$ with $\omega^2 = \sqrt{L_1^2 + L_2^2} - L_1 >0$ and $\chi^2 = \sqrt{L_1^2 + L_2^2} + L_1 >0$. 
With this notation, we write $\hat{\kappa} = \chi + i\omega$. 

\begin{figure}
	\centering
    \includegraphics[width=\textwidth]{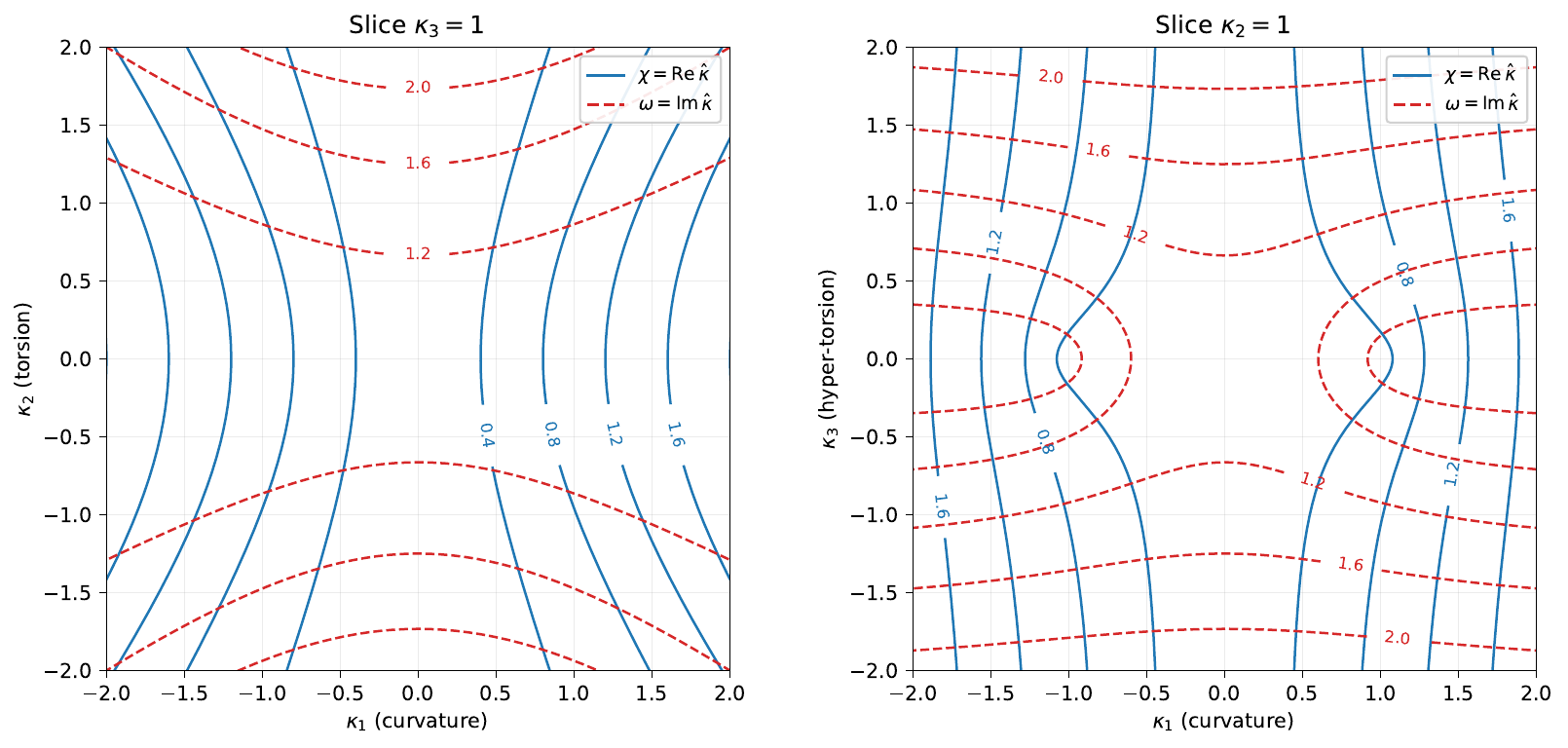}
	\caption{Real ($\chi$, solid blue) and imaginary ($\omega$, dashed red) parts of the complex Frenet-Serret invariant $\hat{\kappa}$ shown over two slices of curvature parameter space: $\kappa_3 = 1$ in the $(\kappa_1,\kappa_2)$-plane (left) and $\kappa_2 = 1$ in the $(\kappa_1,\kappa_3)$-plane (right). Where $\chi$ dominates the response is predominantly exponential amplitude growth; 
    where $\omega$ dominates the response is predominantly oscillatory. 
    }
	\label{fig_kappahat_landscape}
\end{figure}

In Figure \ref{fig_kappahat_landscape}, the real and imaginary components of $\hat{\kappa} = \chi + i\omega$ are displayed for cross-sections of the curvature parameter space.

The frame is given by 
\begin{align}
    {e_0}^\mu =
    \frac{1}{\chi^2+\omega^2}
    \begin{bmatrix}
        \omega^2 + \kappa_1^2 & 0 & \chi^2 - \kappa_1^2 & 0 \\
        0 & \kappa_1(\omega^2 + \kappa_1^2 - \kappa_2^2)/\chi & 0 & \kappa_1(\chi^2 - \kappa_1^2 + \kappa_2^2)/\omega \\
        \kappa_1\kappa_2 & 0 & -\kappa_1\kappa_2 & 0  \\
        0 & \kappa_1\kappa_2\kappa_3/\chi & 0 & -\kappa_1\kappa_2\kappa_3/\omega 
    \end{bmatrix} 
    \begin{bmatrix}
        \cosh(\chi \tau)\\
        \sinh(\chi \tau)\\
        \cos(\omega \tau)\\
        \sin(\omega \tau)
    \end{bmatrix}
\end{align}
and the path of the observer is thus 
\begin{align}\label{stationary_motion_observer_path}
    z(\tau) = 
    \frac{1}{\chi^2+\omega^2}
    \begin{bmatrix}
        \omega^2 + \kappa_1^2 & 0 & \chi^2 - \kappa_1^2 & 0 \\
        0 & \kappa_1(\omega^2 + \kappa_1^2 - \kappa_2^2)/\chi & 0 & \kappa_1(\chi^2 - \kappa_1^2 + \kappa_2^2)/\omega \\
        \kappa_1\kappa_2 & 0 & -\kappa_1\kappa_2 & 0  \\
        0 & \kappa_1\kappa_2\kappa_3/\chi & 0 & -\kappa_1\kappa_2\kappa_3/\omega 
    \end{bmatrix} 
    \begin{bmatrix}
        \frac{1}{\chi}\sinh(\chi \tau)\\
        \frac{1}{\chi}\cosh(\chi \tau)\\
        \frac{1}{\omega}\sin(\omega \tau)\\
        -\frac{1}{\omega}\cos(\omega \tau)
    \end{bmatrix}.
\end{align}

\section{Impacts of nonlinear observer paths on the Dirac spinor wave field}

Consider the plane wave solutions of the Dirac equation 
\begin{align}
    \psi(x) = u(p)e^{-ip_\mu x^\mu}. 
\end{align}
Here, $p^0 = \pm \sqrt{m^2 + p_1^2 + p_2^2 + p_3^2}$ and 
\begin{align}
    u(p) = 
    \begin{bmatrix}
        \phi\\
        \frac{\sigma_1 p_1 + \sigma_2 p_2 + \sigma_3 p_3}{m + |p^0|}\phi
    \end{bmatrix}
\end{align}
is a complex 4-vector that does not depend on $x$. 

The free-particle solution $\psi(x)$ along the path $z(\tau)$ of the noninertial observer is given by 
\begin{align}
    \psi_z(\tau) = D[\Lambda(\tau)]\psi(z(\tau)). 
\end{align}
In particular, for constant curvature parameters:
\begin{align}
    \begin{bmatrix}
        \cosh\left(\frac{\tau}{2}\hat{\kappa}\right) + \frac{\tilde{\kappa}_3}{\hat{\kappa}}\sinh\left(\frac{\tau}{2}\hat{\kappa}\right) & \frac{\tilde{\kappa}_1}{\hat{\kappa}}\sinh\left(\frac{\tau}{2}\hat{\kappa}\right) \\
        \frac{\tilde{\kappa}_1}{\hat{\kappa}}\sinh\left(\frac{\tau}{2}\hat{\kappa}\right) & \cosh\left(\frac{\tau}{2}\hat{\kappa}\right) - \frac{\tilde{\kappa}_3}{\hat{\kappa}}\sinh\left(\frac{\tau}{2}\hat{\kappa}\right)
    \end{bmatrix}
    \begin{bmatrix}
        \psi^0(z(\tau))\\
        \psi^1(z(\tau))
    \end{bmatrix},
\end{align}
where $z(\tau)$ is given by Eq. (\ref{stationary_motion_observer_path}). 
The phase is given by 
\begin{align}
    \Phi(\tau) = p_\mu z^\mu(\tau)
    &= 
    \tilde{p}_\mu \zeta^\mu(\tau)
\end{align} 
\begin{align}
    \tilde{p} &= 
    \begin{bmatrix}
        p_0 & -p_1 & -p_2 & -p_3 
    \end{bmatrix}
    \begin{bmatrix}
        \omega^2 + \kappa_1^2 & 0 & \chi^2 - \kappa_1^2 & 0 \\
        0 & \kappa_1(\omega^2 + \kappa_1^2 - \kappa_2^2)/\chi & 0 & \kappa_1(\chi^2 - \kappa_1^2 + \kappa_2^2)/\omega \\
        \kappa_1\kappa_2 & 0 & -\kappa_1\kappa_2 & 0  \\
        0 & \kappa_1\kappa_2\kappa_3/\chi & 0 & -\kappa_1\kappa_2\kappa_3/\omega 
    \end{bmatrix} \nonumber\\
    \zeta^\mu(\tau) &= 
    \frac{1}{\chi^2+\omega^2}
    \begin{bmatrix}
        \frac{1}{\chi}\sinh(\chi \tau)\\
        \frac{1}{\chi}\cosh(\chi \tau)\\
        \frac{1}{\omega}\sin(\omega \tau)\\
        -\frac{1}{\omega}\cos(\omega \tau)
    \end{bmatrix}.\nonumber
\end{align}
The first component of the transformed spinor is 
\begin{align}
    \psi_z^0(\tau) 
    &= 
    \frac{1}{2}\left(\left(1 + \frac{\tilde{\kappa}_3}{\hat{\kappa}}\right)\phi^0 + \frac{\tilde{\kappa}_1}{\hat{\kappa}}\phi^1\right)\exp\left(\frac{\tau}{2}\hat{\kappa}-i\tilde{p}_\mu \zeta^\mu(\tau)\right)\nonumber\\
    &+ 
    \frac{1}{2}\left(\left(1 - \frac{\tilde{\kappa}_3}{\hat{\kappa}}\right)\phi^0 - \frac{\tilde{\kappa}_1}{\hat{\kappa}}\phi^1\right)\exp\left(-\frac{\tau}{2}\hat{\kappa}-i\tilde{p}_\mu \zeta^\mu(\tau)\right). 
\end{align}
Since $\hat{\kappa}$ is complex, the spinor transformation contributes to phase aberrations. 
The phase and amplitude are 
\begin{align}
    \Phi(\tau) &= \frac{\tau}{2}\omega - \tilde{p}_\mu \zeta^\mu(\tau) \label{stationary_phase}\\
    \mathcal{A}(\tau) &= \frac{1}{2}\left(\left(1 + \frac{\tilde{\kappa}_3}{\hat{\kappa}}\right)\phi^0 + \frac{\tilde{\kappa}_1}{\hat{\kappa}}\phi^1\right)\exp\left(\frac{\tau}{2}\chi\right). \label{stationary_amplitude}
\end{align}

\begin{figure}
	\centering
    \includegraphics[width=0.7\textwidth]{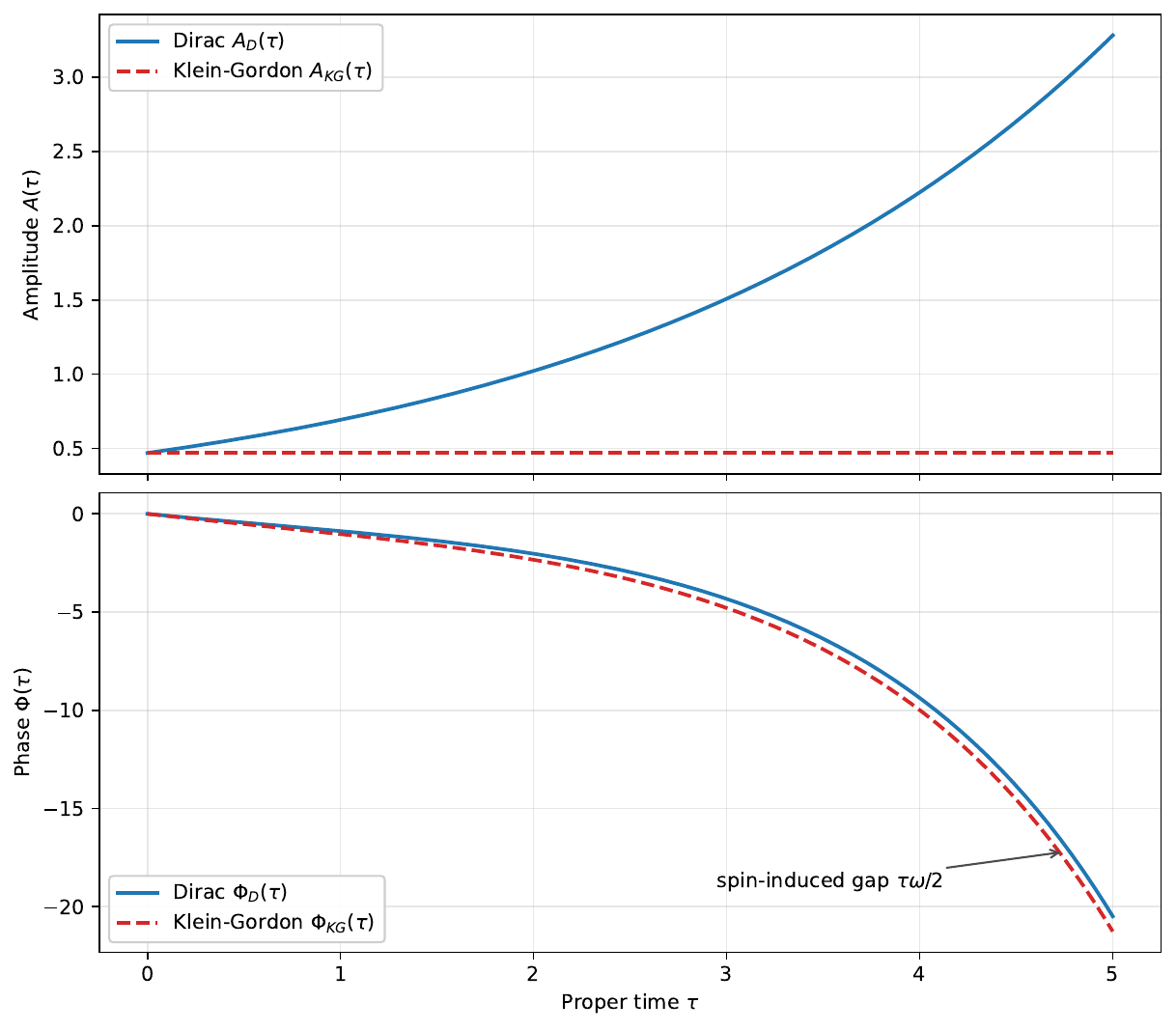}
	\caption{Dirac (solid blue) and Klein-Gordon (dashed red) amplitude (top) and phase (bottom) along the observer path with constant curvature parameters $\kappa_1 = 0.8$, $\kappa_2 = 0.2$, $\kappa_3 = 0.3$, giving rapidity $\chi = 0.78$ and rotation frequency $\omega = 0.31$. The initial 2-spinor is $\phi = (1,0)^T$ and the Klein-Gordon amplitude is normalized to match the Dirac amplitude at $\tau=0$. The widening gap between the phase curves at large $\tau$ is precisely the spin-induced contribution $\tau\omega/2$ identified in Equation (\ref{stationary_phase}) below. Curves are plotted in natural units ($c=\hbar=1$) with the particle mass set to $m=1$. All other quantities are in units of $m$ or $1/m$.}
	\label{fig_dirac_vs_KG}
\end{figure}
For comparison, the scalar Klein-Gordon field along the same worldline carries only the kinematic phase $\Phi_{\rm KG}(\tau) = -p_\mu z^\mu(\tau)$ with constant amplitude $\mathcal{A}_{\rm KG}(\tau) = \mathcal{A}_{\rm KG}(0)$.
In Figure \ref{fig_dirac_vs_KG}, the phase and amplitude of the Dirac and Klein-Gordon fields are compared. The spin of the Dirac field produces a continuous phase shift as a function of proper time and the amplitude, which is constant in the Klein-Gordon field, grows exponentially for the Dirac field. 
The vertical gap between the two phase curves is precisely the spin-induced $\tau\omega/2$ contribution introduced by the complex $\hat{\kappa}$.

For variable acceleration, e.g., constant proper jolt in the $x^1$ direction, the observer trajectory is 
\begin{align}
    z(\tau) = 
    \begin{bmatrix}
    \int_0^{\tau} \cosh(\chi(\tau'))\, d\tau' \\
    \int_0^{\tau} \sinh(\chi(\tau'))\, d\tau' \\
    0 \\
    0 
    \end{bmatrix}
\end{align}
where $\chi(\tau) = \chi_0 + a_0\tau + j_0 (\tau)^2/2$. 
The free-particle solution $\psi(x)$ along the path $z(\tau)$ is given by 
\begin{align}
    \begin{bmatrix}
        \cosh(\chi(\tau)/2) & -\sinh(\chi(\tau)/2) \\
        -\sinh(\chi(\tau)/2) & \cosh(\chi(\tau)/2) 
    \end{bmatrix}
    \begin{bmatrix}
        \psi^0(z(\tau))\\
        \psi^1(z(\tau))
    \end{bmatrix}.
\end{align}
The first component of the transformed spinor is 
\begin{align}
    \psi_z^0(\tau) 
    = (\cosh(\chi(\tau)/2)\phi^0 &- \sinh(\chi(\tau)/2)\phi^1)\\
    &\times\exp(-i(p_0\int_0^{\tau} \cosh(\chi(\tau'))\, d\tau' - p_1\int_0^{\tau} \sinh(\chi(\tau'))\, d\tau'))\nonumber .
\end{align}
The phase and amplitude are 
\begin{align}
    \Phi(\tau) &= -p_0\int_0^{\tau} \cosh(\chi(\tau'))\, d\tau' + p_1\int_0^{\tau} \sinh(\chi(\tau'))\, d\tau'\label{jolt_phase}\\
    \mathcal{A}(\tau) &= \cosh(\chi(\tau)/2)\phi^0 - \sinh(\chi(\tau)/2)\phi^1. \label{jolt_amplitude}
\end{align}

Equations (\ref{jolt_phase}) and (\ref{jolt_amplitude}) extend our framework beyond stationary worldlines into the regime of non-stationary kinematics, where three features distinguish the constant-proper-jolt response from the Rindler and constant-curvature-parameter cases analyzed above. 
The rapidity grows quadratically in proper time, so the amplitude factor scales as $\exp(j_0 \tau^2 / 4)$ at large $\tau$ which is a super-exponential response which is distinct from the exponential growth obtained from any stationary worldline. 
The amplitude itself depends sensitively on the initial 2-spinor: generic states grow monotonically through the $\cosh$ contribution, but on the marginal state $\phi^0 = \phi^1$ the two terms cancel to leave $\exp(-\chi(\tau)/2)\phi^0$, a Dirac configuration that decays under acceleration with no analogue in the Klein-Gordon field. 
The phase integrals in (\ref{jolt_phase}) reduce to Fresnel-type functions rather than elementary expressions, encoding a continuous chirp in the instantaneous frequency that is the direct quantum-spinor analogue of the jolt-induced classical Doppler chirp identified for nonlinear sensor motion in \cite{barclay2024doppler,barclay2024physics,barclay2026spectralasilomar,barclay2026dynamic}. 
The constant-jolt regime is also the physically relevant one for any experimental platform in which the proper acceleration is not strictly constant — including electrons in finite-envelope ultraintense laser pulses \cite{chen1999testing,schutzhold2006signatures} and the ramp-up phase of accelerator beams \cite{sheehy2024applications}. 
Thus, Equations (\ref{jolt_phase}) and (\ref{jolt_amplitude}) provide the appropriate spinor templates for predictions in those settings.
 
\section{Application to $\beta^+$ decay}\label{sec:beta-plus-decay}

The process of a proton $p^+$ decaying into a neutron $n^0$, positron $e^+$, and neutrino $\nu$ is called $\beta^+$ decay or positron emission. 
In \cite{vanzella2001decay,suzuki2003analytic}, it was shown that, in the framework of standard quantum field theory, a $\beta^+$ decay process is induced by the acceleration of a proton. 
In this section, we recall how acceleration induces proton decay and extend this result to arbitrary rapidity, and, in particular, constant proper jolt. 

\subsection{Constant proper acceleration}
We calculate the transition amplitude $\mathcal{A}$ for this process to first order.
For this calculation, we use the Gamma matrices in the Dirac representation:
\begin{align}
    \gamma^0 = 
    \begin{bmatrix} 
        I_2 & 0 \\ 
        0 & -I_2 
    \end{bmatrix}, 
    \qquad
    \gamma^i = 
    \begin{bmatrix} 
        0 & \sigma^i \\ 
        -\sigma^i & 0 
    \end{bmatrix}.
\end{align}

The Lagrangian density is
\begin{align}
    \mathcal{L} = \overline{\Psi_{e}}(i\gamma^{\mu}\partial_{\mu}- m_{e})\Psi_{e} + \overline{\Psi_{\nu}}(i\gamma^{\mu}\partial_{\mu}- m_{\nu})\Psi_{\nu} + \mathcal{L}_{h} + j_{\mu}(\overline{\Psi_{\nu}}\gamma^\mu\Psi_{e} + \overline{\Psi_{e}}\gamma^\mu\Psi_{\nu})
\end{align}
where $\mathcal{L}_{h}$ is the effective Lagrangian of the hadrons and $\overline{\Psi} = \Psi^{\dagger}\gamma^0$. 
The Fermi action is 
\begin{align}
    S_{I} = \int d^2x \,j_{\mu} [\overline{\Psi}_{\nu}\gamma^\mu \Psi_{e} + \overline{\Psi}_{e}\gamma^\mu \Psi_{\nu}].
\end{align}

The initial and final states are non-interacting states (eigenstates of the free theory). 
The initial state is the Minkowski vacuum with the semiclassical current in the proton state:
\begin{align}
    |i\rangle = |p^+\rangle\otimes|0_{M}\rangle
\end{align}
The final state is a positron and neutrino together with the semiclassical current in the neutron state:
\begin{align}
    |f\rangle = |n^0\rangle\otimes d_{p',\sigma'}^{e\dagger}b_{p,\sigma}^{\nu\dagger}|0_{M}\rangle
\end{align}

To first order, the quantum amplitude is 
\begin{align}
    \mathcal{A} = \langle f | \left(1 - i\int\,dt'H_{I}(t')\right) | i \rangle
    &= 
    i\langle f | \int\int\,dt' dx j_{\mu}(\overline{\Psi_{\nu}}\gamma^\mu\Psi_{e} + \overline{\Psi_{e}}\gamma^\mu\Psi_{\nu}) | i \rangle\\
    &= 
    i\int\int\,dt' dx \langle n^0| j_{\mu} |p^+\rangle \langle 0_M  |  b^\nu_{p,\sigma}  d^e_{p',\sigma'}(\overline{\Psi_{\nu}}\gamma^\mu\Psi_{e} + \overline{\Psi_{e}}\gamma^\mu\Psi_{\nu}) |  0_M \rangle\nonumber 
\end{align}

The quantum fields are operators in the Heisenberg picture of the free-field theory:
\begin{align}
    \Psi_{e}(x) &= \sum_{\sigma=\pm}\int \frac{d\mathbf{p}}{\sqrt{ 2\pi }}\,[b_{\mathbf{p},\sigma}^e\mathcal{U}_{\sigma}^e(x;p) + d_{\mathbf{p},\sigma}^{e\dagger}\mathcal{U}_{-\sigma}^e(x;-p)]\\
    \Psi_{\nu}(x) &= \sum_{\sigma=\pm}\int \frac{d\mathbf{p}}{\sqrt{ 2\pi }}\,[b_{\mathbf{p},\sigma}^{\nu}\mathcal{U}_{\sigma}^\nu(x;p) + d_{\mathbf{p},\sigma}^{\nu\dagger}\mathcal{U}_{-\sigma}^\nu(x;-p)]\\
    j^\mu(x) &= q(a^{-1}\sinh^{-1}(at))u^\mu \delta(\sqrt{ z^2-t^2 }-a^{-1})
\end{align}
where 
\begin{align}
    u^\mu &= \left(\sqrt{ (at)^2+1 },at\right) = \left(\cosh(a\tau), \sinh(a\tau)\right),\\
    q(\tau) &= e^{iH_{h}\tau}q_{0}e^{-iH_{h}\tau}.
\end{align}
The operator $q_{0}$ acts on the two-level quantum system (i.e., 2-D Hilbert space) with (Schrödinger-picture) Hamiltonian $H_{h}$ satisfying $H_{h}|n^0\rangle = m_{n}|n^0\rangle$ and $H_{h}|p^+\rangle = m_{p}|p^+\rangle$. 
The operator $q_{0}$ provides the dimensionless effective Fermi constant $G_{F}=|\langle m_{p}|q_{0}|m_{n}\rangle|$. 

We have the following functions
\begin{align}
    \mathcal{U}_{+}(t,z;\pm p_{0},p_{3})
    &=
    \exp(i(\mp p_{0}t+p_{3}z))
    \begin{bmatrix}\pm \sqrt{ \frac{p_{0}\pm m}{2p_{0}} } & 0 & \frac{p_{3}}{\sqrt{ 2p_{0} (p_{0} \pm m)}} & 0\end{bmatrix}^T\\
    \mathcal{U}_{-}(t,z;\pm p_{0},p_{3})
    &=
    \exp(i(\mp p_{0}t+p_{3}z))
    \begin{bmatrix}0 & \pm \sqrt{ \frac{p_{0}\pm m}{2p_{0}} } & 0 & \frac{-p_{3}}{\sqrt{ 2p_{0} (p_{0} \pm m)}}\end{bmatrix}^T.
\end{align}

The anti-commutation relations for the annihilation/creation operators are
\begin{align}
    \{b_{p,\sigma}^f, b^{f'\dagger}_{p',\sigma'}\} &= \{d_{p,\sigma}^f, d^{f'\dagger}_{p',\sigma'}\} = \delta^{ff'}\delta(p-p')\delta_{\sigma\sigma'}, \\
    0 &= \{b, b\} = \{b, d\} = \{b, d^\dagger\} = \{b^\dagger, d\} = \{b^\dagger, d^\dagger\} = \{d, d\}.
\end{align}
In particular: 
\begin{equation}
     \{b_{p,\sigma}, b^{\dagger}_{p',\sigma'}\} = \{d_{p,\sigma}, d^{\dagger}_{p',\sigma'}\} = \delta(p-p') \delta_{\sigma\sigma'} 
\end{equation}
\begin{equation}
    \{d^{e},  b^{\nu\dagger}\}  =  0.
\end{equation}

The computation of the transition amplitude $\mathcal{A}$ requires several steps (see Appendix~\ref{sec:beta-plus-decay-appendix}). The final form is 
\begin{align}\label{eq:A_final_accel}
    \mathcal{A} 
    = 
    \frac{-G_F\delta_{\sigma',-\sigma}}{4\pi\sqrt{p_0p_0'(p_0+m_\nu)(p_0'-m_e)}} \int d\tau\, e^{i\Phi(\tau)} \bigg\{ &\big[(p_0+m_\nu)(p_0'-m_e) + p_3 p_3'\big]\cosh(a\tau)\nonumber\\
    - &\big[(p_0+m_\nu)p_3' + (p_0'-m_e)p_3\big]\sinh(a\tau)\bigg\},
\end{align}
with $\Phi(\tau) = \Delta m\tau + a^{-1}(p_0'+p_0)\sinh(a\tau) - a^{-1}(p_3'+p_3)\cosh(a\tau)$. Only the opposite-spin ($\sigma' = -\sigma$) contributes. 

\subsection{Constant proper jolt}

For a proton undergoing initial proper acceleration $a_0$ and constant proper jolt $j_0$, the rapidity is quadratic in proper time:
\begin{align}
    \chi(\tau) = \chi_0 + a_0\tau + \frac{1}{2} j_0\tau^2,
\end{align}
with $\chi_0$ fixed by the initial velocity via $\tanh(\chi_0) = v_0/c$.  
The four-velocity is
\begin{align}
    u^\mu(\tau) = \big(\cosh(\chi(\tau)),  \sinh(\chi(\tau))\big),
\end{align}
and the worldline follows by integration:
\begin{align}
    t(\tau) = \int_0^\tau \cosh(\chi(\tau'))\,d\tau', 
    \qquad
    z(\tau) = \int_0^\tau \sinh(\chi(\tau'))\,d\tau'.
\end{align}
Setting $j_0 = 0$ recovers the constant-acceleration case with $a = a_0$.

The vacuum matrix element of the lepton bilinear is
\begin{align}
    \langle 0_M| b^\nu_{p,\sigma} d^e_{p',\sigma'} \bar\Psi_\nu \gamma^\mu \Psi_e |0_M \rangle
    =  
    -\frac{1}{2\pi} \mathcal{U}^{\nu\dagger}_\sigma(x;p) \gamma^0\gamma^\mu \mathcal{U}^e_{-\sigma'}(x;-p').
\end{align}
Substituting this together with $\langle n^0|q(\tau)|p^+\rangle = G_F e^{i\Delta m\tau}$ gives
\begin{align}\label{eq:A_general}
    \mathcal{A} = -\frac{G_F}{2\pi}\int d\tau e^{i\Delta m\tau} u_\mu(\tau) \widetilde{\mathcal{U}}^{\nu\dagger}_\sigma(p) \gamma^0\gamma^\mu \widetilde{\mathcal{U}}^e_{-\sigma'}(-p') E_j(\tau),
\end{align}
where the phase evaluated on the constant-jolt worldline is
\begin{align}
    E_j(\tau)  
    &= 
    \exp\Big[i(p_0+p_0')t(\tau) - i(p_3+p_3')z(\tau)\Big]\nonumber\\
    &= 
    \exp\left[i\int_0^\tau\Big((p_0+p_0')\cosh(\chi(\tau')) - (p_3+p_3')\sinh(\chi(\tau'))\Big) d\tau'\right].
\end{align}

Similar to the constant-acceleration case: 
\begin{align}
    \widetilde{\mathcal{U}}^{\nu\dagger}_\sigma(p)\gamma^0\gamma^0\widetilde{\mathcal{U}}^e_{-\sigma'}(-p') 
    = 
    \delta_{\sigma,-\sigma'}\mathcal{C}_0, 
    \qquad
    \widetilde{\mathcal{U}}^{\nu\dagger}_\sigma(p)\gamma^0\gamma^1\widetilde{\mathcal{U}}^e_{-\sigma'}(-p') 
    = 
    \delta_{\sigma,-\sigma'}\mathcal{C}_1,
\end{align}
with scalar coefficients
\begin{align}
    \mathcal{C}_0 
    &= 
    -\frac{(p_0+m_\nu)(p_0'-m_e) + p_3p_3'}{2\sqrt{p_0p_0'(p_0+m_\nu)(p_0'-m_e)}},\\
    \mathcal{C}_1 
    &= 
    -\frac{(p_0+m_\nu)p_3' + (p_0'-m_e)p_3}{2\sqrt{p_0p_0'(p_0+m_\nu)(p_0'-m_e)}}.
\end{align}
Thus:
\begin{align}\label{eq:u_contraction}
    u^\mu\widetilde{\mathcal{U}}^{\nu\dagger}_\sigma(p)\gamma^0\gamma_\mu\widetilde{\mathcal{U}}^e_{-\sigma'}(-p') 
    = 
    \delta_{\sigma,-\sigma'}\big[\mathcal{C}_0\cosh(\chi(\tau)) - \mathcal{C}_1\sinh(\chi(\tau))\big].
\end{align}

Substituting \eqref{eq:u_contraction} and the explicit $\mathcal{C}_{0,1}$ into \eqref{eq:A_general} gives the single amplitude
\begin{align}\label{eq:A_jolt}
    \mathcal{A} = \frac{-G_F \delta_{\sigma',-\sigma}}{4\pi\sqrt{p_0p_0'(p_0 + m_\nu)(p_0' - m_e)}}\int d\tau\, e^{i\Phi_j(\tau)}B(\tau),
\end{align}
where 
\begin{eqnarray}
    B(\tau) = ((p_0 + m_\nu)(p_0' - m_e) + p_3p_3')\cosh\chi(\tau) - ((p_0 + m_\nu) p_3' + (p_0' - m_e) p_3)\sinh\chi(\tau) \\
    \Phi_j(\tau) = \Delta m\tau + (p_0+p_0')\int_0^\tau \cosh(\chi(\tau'))\,d\tau' - (p_3+p_3')\int_0^\tau \sinh(\chi(\tau'))\,d\tau'.\label{eq:Phi_j}
\end{eqnarray}
This calculation of the quantum transition amplitude (\ref{eq:A_jolt}) extends the estimates of noninertial proton decay \cite{vanzella2001decay,suzuki2003analytic} to nonlinear rapidity. 

\begin{figure}
	\centering
    \includegraphics[width=\textwidth]{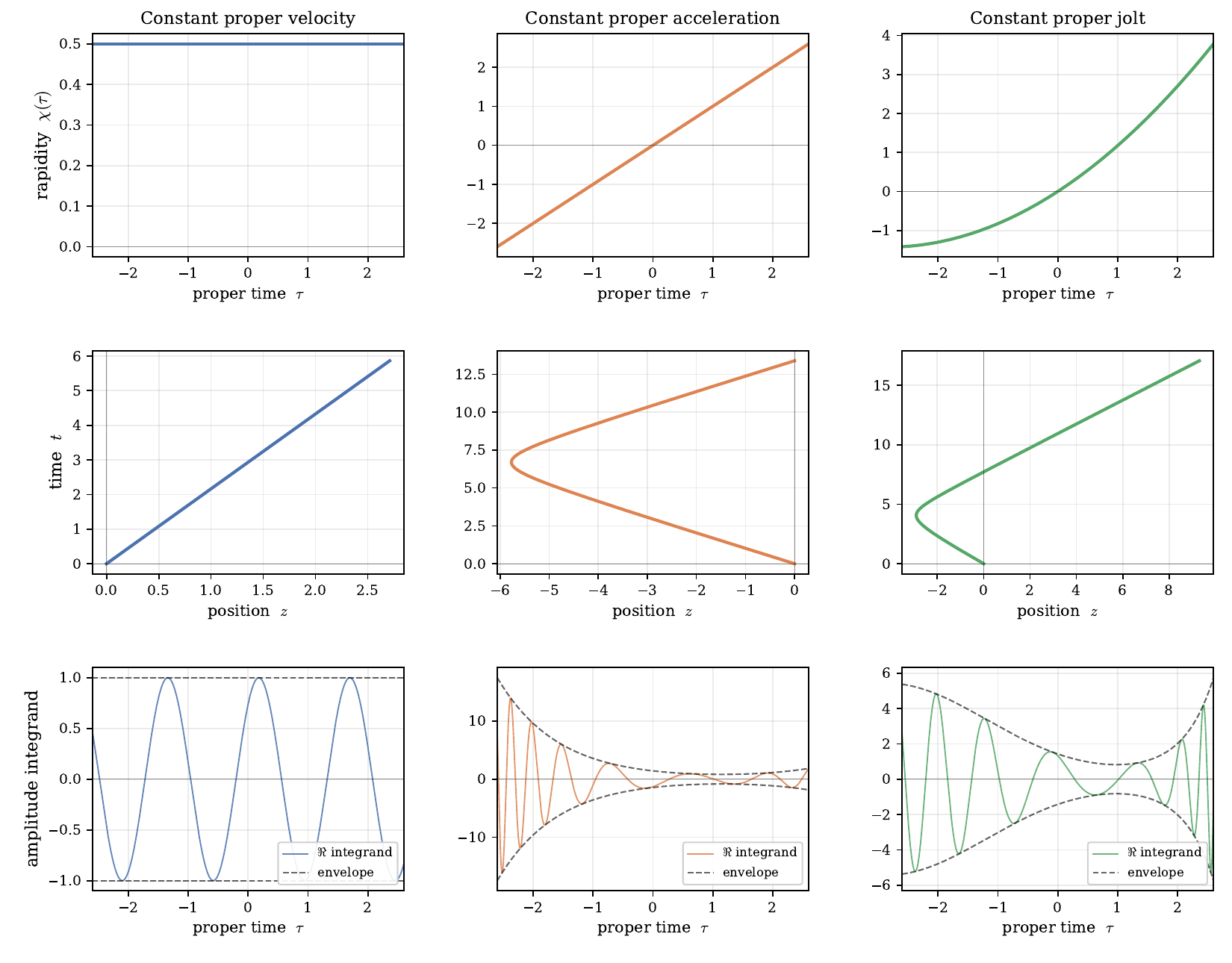}
	\caption{Comparison of the quantum transition amplitude integrand (bottom row) for $\beta^+$ decay of a proton with constant velocity (column 1), constant proper acceleration (column 2), and constant proper jolt (column 3). Curves are plotted in natural units ($c=\hbar=1$) with $a_0 = 1$, $j_0/a_0^{2} = 0.35$, $\Delta m/a_0 = 2$, and lepton momenta $p_0, p_3, p_0', p_3' \sim a_0$. For large proper time $\tau$, the amplitude integrand is asymptotic to (bottom center) $\exp(a_0|\tau|)$ and (bottom right) $\exp(j_0\tau^2/2)$.}
	\label{fig_beta_plus}
\end{figure}

Figure~\ref{fig_beta_plus} shows how the proper-time integrand for the transition amplitude changes across three successive source-proton kinematic regimes: uniform proper velocity, uniform proper acceleration, and uniform proper jolt. 
For inertial motion, $\chi$ remains fixed, and the trajectory is a timelike straight line, so the integrand oscillates under a bounded envelope, as expected for the stability of a free proton in the standard model. 
Under constant proper acceleration, the trajectory becomes a Rindler hyperbola and the envelope grows as $\exp(a_0|\tau|)$. 
This is the result described in \cite{vanzella2001decay,matsas1999decay}. 
In the constant-jolt generalization developed here, the quadratic rapidity $\chi(\tau)=\chi_0+a_0\tau+\tfrac{1}{2}j_0\tau^2$ produces a super-hyperbolic trajectory and a super-exponentially growing envelope, $\exp(j_0\tau^2/2)$. 
The altered large-$|\tau|$ behavior provides the formal distinction between the jolt-corrected result and constant-acceleration. 
Generalizing predictions of noninertial motion in quantum field theory to nonlinear rapidity creates a useful computational framework for Unruh-type predictions of arbitrary motion.  

\subsection{Instantaneous transition rate under constant proper jolt}

We now obtain a decay rate for the proton undergoing constant proper jolt. 
The differential transition rate is calculated by squaring the amplitude and summing over spin configurations: 
\begin{align}
    \frac{\partial^2 P}{\partial p_3 \partial p_3'} 
    = 
    \sum_{\sigma,\sigma'} |\mathcal{A}|^2 
    = 
    \frac{G_F^2}{8\pi^2 p_0p_0'(p_0 + m_\nu)(p_0' - m_e)}
    \int\int d\tau_1\,d\tau_2\, e^{i[\Phi_j(\tau_1)-\Phi_j(\tau_2)]}B(\tau_1)B(\tau_2).
\end{align}
Next, we transform the integration variables to a sum and difference: 
\begin{align}
    s = \frac{1}{2}(\tau_1 + \tau_2),\quad \xi = \tau_1 - \tau_2. 
\end{align}
We define the lepton-pair invariant mass $M$ and its rapidity $\Theta$:
\begin{align}
    p_0+p_0' = M\cosh(\Theta), 
    \qquad 
    p_3+p_3' = M\sinh(\Theta),
    \qquad 
    M = \sqrt{(p_0+p_0')^2 - (p_3+p_3')^2},
\end{align}
so that
\begin{align}
    (p_0+p_0')\cosh(\chi) - (p_3+p_3')\sinh(\chi) = M\cosh(\chi-\Theta)
\end{align}
for any rapidity $\chi$.
The phase difference is 
\begin{align}
    \Delta\Phi(s,\xi) 
    = 
    \Phi_j(\tau_1) - \Phi_j(\tau_2) 
    &= 
    \Delta m \xi + M\int_{\tau_2}^{\tau_1} \cosh\left(\chi(\tau')-\Theta\right)\,d\tau' \nonumber\\
    &= 
    \Delta m \xi + M\int_{-\xi/2}^{\xi/2} \cosh\left[-\tilde{\Theta}(s) + \alpha(s)u + \frac{1}{2}j_0u^2\right]\,du,
\end{align}
where $\tilde{\Theta}(s) = \Theta - \chi(s)$ and $\alpha(s) = a_0 + j_0 s$.

Expanding about $s$ gives
\begin{align}
    \chi(s+u) = \chi(s) + \alpha(s)u + \frac{1}{2}j_0u^2.
\end{align}
We have 
\begin{align}
    \chi(\tau_1) + \chi(\tau_2) = 2\chi(s) + \frac{1}{4}j_0\xi^2,
    \qquad
    \chi(\tau_1) - \chi(\tau_2) = \alpha(s)\xi
\end{align}
and 
\begin{align}
    \frac{B(\tau_1)B(\tau_2)}{2(p_0 + m_\nu)(p_0' - m_e)}
    = 
    \left(p_0p_0' + p_3p_3'\right)\cosh(\Xi(s,\xi)) 
    &- \left(p_0p_3' + p_0'p_3\right)\sinh(\Xi(s,\xi)) \nonumber\\
    &- m_\nu m_e\cosh\left(\alpha(s)\xi\right),
\end{align}
where we define the jolt-shifted sum rapidity: 
\begin{align}
    \Xi(s,\xi) = 2\chi(s) + \frac{1}{4}j_0\xi^2.
\end{align}
The differential transition rate is thus
\begin{align}
    \frac{d^2P}{dp_3dp_3'}
    = 
    \frac{G_F^2}{4\pi^2p_0p_0'}\int_{-\infty}^{\infty}ds\,\int_{-\infty}^{\infty}d\xi\,e^{i\Delta\Phi(s,\xi)}\bigg[&\left(p_0p_0' + p_3p_3'\right)\cosh(\Xi(s,\xi))\nonumber\\
    -&
    \left(p_0p_3' + p_0'p_3\right)\sinh(\Xi(s,\xi))\nonumber\\
    -&
    m_\nu m_e\cosh(\alpha(s)\xi)\bigg].
\end{align}

In contrast to the constant acceleration case, we cannot factor out $\int\,ds$ to obtain the constant decay rate. Therefore, we calculate the instantaneous transition rate $\dot{P}$ using the relation:
\begin{align}
    \frac{d^2P}{dp_3dp_3'}
    = 
    \int_{-\infty}^\infty ds \frac{d^2\dot{P}(s)}{dp_3dp_3'}
\end{align} 
giving 
\begin{align}
    \dot{P}(s) 
    = 
    \frac{G_F^2}{4\pi^2p_0p_0'}\int_{-\infty}^{\infty}dp_3\int_{-\infty}^{\infty}dp_3'\,\int_{-\infty}^{\infty}d\xi\,e^{i\Delta\Phi(s,\xi)}\bigg[&\left(p_0p_0' + p_3p_3'\right)\cosh(\Xi(s,\xi))\nonumber\\
    -&
    \left(p_0p_3' + p_0'p_3\right)\sinh(\Xi(s,\xi))\nonumber\\
    -&
    m_\nu m_e\cosh(\alpha(s)\xi)\bigg].
\end{align}
Next, we boost by rapidity $-\chi(s)$ (rather than $-as$) 
\begin{align}
    \tilde{p}_3 = -p_0\sinh(\chi(s)) + p_3\cosh(\chi(s)),
    \qquad
    \tilde{p}_3' = -p_0'\sinh(\chi(s)) + p_3'\cosh(\chi(s)),
\end{align}
with $\tilde{p}_0 = \sqrt{\tilde{p}_3^2 + m_\nu^2}$ and $\tilde{p}_0' = \sqrt{\tilde{p}_3'^2 + m_e^2}$.
The instantaneous differential transition rate becomes 
\begin{align}
    \frac{d^2\dot P(s)}{d\tilde{p}_3d\tilde{p}_3'}
    = 
    \frac{G_F^2}{4\pi^2\tilde{p}_0\tilde{p}_0'}\int_{-\infty}^{\infty}d\xi\,
    e^{i\Delta\Phi(s,\xi)}
    \bigg[&\cosh\left(\frac{j_0\xi^2}{4}\right)\left(\tilde{p}_0\tilde{p}_0' + \tilde{p}_3\tilde{p}_3'\right)\nonumber\\
    -&
    \sinh\left(\frac{j_0\xi^2}{4}\right)\left(\tilde{p}_0\tilde{p}_3' + \tilde{p}_0'\tilde{p}_3\right)\nonumber\\
    -&
    m_\nu m_e\cosh\left(\alpha(s)\xi\right)\bigg],
\end{align}
where
\begin{align}
    \Delta\Phi(s,\xi) 
    = 
    \Delta m\xi + M\int_{-\xi/2}^{\xi/2}\cosh\left[-\tilde{\Theta}(s) + \alpha(s)u + \frac{1}{2}j_0u^2\right] du.
\end{align}
The decay rate is 
\begin{align}
    \Gamma_{\text{jolt}}(s) 
    = 
    \dot P(s) 
    = 
    \int d\tilde{p}_3 \int d\tilde{p}_3'\frac{d^2\dot P(s)}{d\tilde{p}_3d\tilde{p}_3'}.
\end{align}

To compare to \cite{matsas1999decay}, we set $j_0= 0$, giving 
\begin{align}
    \dot P(s) 
    = 
    \frac{G_F^2}{4\pi^2\tilde{p}_0\tilde{p}_0'}
    \int d\tilde{p}_3 \int d\tilde{p}_3'\int d\xi\,e^{i\Delta\Phi(s,\xi)}
    \bigg[\tilde{p}_0\tilde{p}_0' + \tilde{p}_3\tilde{p}_3' - m_\nu m_e\cosh\left(a_0\xi\right)\bigg]
\end{align}
and 
\begin{align}
    \Delta\Phi 
    &= 
    \Delta m \xi + M\int_{-\xi/2}^{\xi/2} \cosh\left[-\tilde{\Theta}(s) + a_0u\right]\,du\\
    &= 
    \Delta m \xi + \frac{M}{a}\left[\sinh\left(-\tilde{\Theta} + \frac{a\xi}{2}\right) + \sinh\left(\tilde{\Theta} + \frac{a\xi}{2}\right)\right]
    = 
    \Delta m \xi + \frac{2M}{a}\cosh(\tilde{\Theta})\sinh\left(\frac{a\xi}{2}\right),\nonumber 
\end{align}
with $M\cosh(\tilde{\Theta}(s)) = M\cosh(as - \Theta) = (p_0 + p_0')\cosh(as) - (p_3 + p_3')\sinh(as) = \tilde{p}_0 + \tilde{p}_0'$.
Thus, the instantaneous transition rate $\dot P$ reduces to the proton decay rate $\Gamma_{\mathrm{accel}}$ derived in \cite{matsas1999decay}. The additional dependence on proper jolt appears as the integral of $\exp(j_0\xi^2/4)$. For a noninertial proton, the decay rate is determined locally in proper time: it admits an expansion in proper acceleration and its derivatives, rather than an expression as a functional of the full past trajectory $z^\mu(\tau)$.

\section{Conclusion}

We derived the dynamic Doppler effects of noninertial observer motion on Dirac spinor fields and obtained, in closed form, the transformed free-particle wave function for an arbitrary stationary worldline characterized by curvature, torsion, and hyper-torsion, together with time-dependent proper acceleration and proper jolt. 
The central observation is that the Frenet-Serret invariant $\hat{\kappa}$ entering the spinor representation is complex; its real part $\chi$ controls the amplitude response while its imaginary part $\omega$ enters the phase as a spin-induced contribution $\tau\omega/2$ with no analogue in the scalar Klein-Gordon case. 
The same trajectory therefore produces a two-frequency Dirac spectrum in contrast to the scalar field which has a single frequency. 
These spinor corrections set the prediction against which any laser-driven, condensed-matter analog, or Unruh-DeWitt-detector experiment performed on a fermionic system should be compared, and they sharpen what was previously a qualitative analogy between noninertial Dirac dynamics and the dynamic Doppler effects of classical electromagnetism into a quantitative framework.
Our derivation of the proper-jolt effects on quantized field-theoretic $\beta^+$ decay provides a general framework for testable variable-rapidity Unruh predictions.

\appendix

\section{Quantum transition amplitude for $\beta^+$-decay}\label{sec:beta-plus-decay-appendix}

Calculation of the amplitude $\mathcal{A}$ obtained in Section~\ref{sec:beta-plus-decay} requires several steps. Here we detail the constant proper acceleration case. 
First, calculate
\begin{align}
    \langle n^0|q(\tau)|p^+\rangle 
    = 
    \langle n^0|e^{iH_{h}\tau}q_{0}e^{-iH_{h}\tau}|p^+\rangle 
    = 
    e^{i(m_{n}-m_{p})\tau}\langle n^0|q_{0}|p^+\rangle 
    = 
    e^{i(m_{n}-m_{p})\tau}G_{F}
\end{align}
and transform the $\delta$ function in the definition of $j^\mu$:
\begin{align}
    \delta(\sqrt{ z^2-t^2 }-a^{-1}) = \frac{\delta(z-\sqrt{ t^2+a^{-2} })}{az}.
\end{align}

Next, calculate
\begin{align}
    &\langle 0_M | b^\nu_{p,\sigma} d^e_{p',\sigma'} \bar\Psi_\nu \gamma^\mu \Psi_e | 0_M \rangle\\
    &\quad = \sum_{\sigma_e=\pm}\sum_{\sigma_\nu=\pm} \int \frac{dp_e}{\sqrt{2\pi}}\int \frac{dp_\nu}{\sqrt{2\pi}} \langle 0_M | b^\nu_{p,\sigma} d^e_{p',\sigma'} \Big[b^{\nu\dagger}_{p_\nu,\sigma_\nu} \mathcal{U}^{\nu\dagger}_{\sigma_\nu}(x;p_\nu) + d^{\nu}_{p_\nu,\sigma_\nu} \mathcal{U}^{\nu\dagger}_{-\sigma_\nu}(x;-p_\nu) \Big] \nonumber\\
    &\qquad \cdot \gamma^0\gamma^\mu \Big[b^e_{p_e,\sigma_e} \mathcal{U}^e_{\sigma_e}(x;p_e) + d^{e\dagger}_{p_e,\sigma_e} \mathcal{U}^e_{-\sigma_e}(x;-p_e) \Big] | 0_M \rangle \\
    &\quad = \sum_{\sigma_e=\pm}\sum_{\sigma_\nu=\pm} \int \frac{dp_e}{\sqrt{2\pi}}\int \frac{dp_\nu}{\sqrt{2\pi}} \mathcal{U}^{\nu\dagger}_{\sigma_\nu}(x;p_\nu) \gamma^0\gamma^\mu \mathcal{U}^e_{-\sigma_e}(x;-p_e) \langle 0_M | b^\nu_{p,\sigma} d^e_{p',\sigma'} b^{\nu\dagger}_{p_\nu,\sigma_\nu} d^{e\dagger}_{p_e,\sigma_e} | 0_M \rangle,
\end{align}
where
\begin{align}
    &\langle 0_M | b^\nu_{p,\sigma} d^e_{p',\sigma'} b^{\nu\dagger}_{p_\nu,\sigma_\nu} d^{e\dagger}_{p_e,\sigma_e} | 0_M \rangle 
    = -\delta(p_e - p') \delta_{\sigma_e\sigma'} \delta(p_\nu - p) \delta_{\sigma_\nu\sigma}.
\end{align}
Therefore,
\begin{align}
    \langle 0_M | b^\nu_{p,\sigma} d^e_{p',\sigma'} \bar\Psi_\nu \gamma^\mu \Psi_e | 0_M \rangle
    =  
    -\frac{1}{2\pi} \mathcal{U}^{\nu\dagger}_{\sigma}(x;p) \gamma^0\gamma^\mu \mathcal{U}^e_{-\sigma'}(x;-p').
\end{align}
The scalar factors of $\mathcal{U}^{\nu\dagger}$ and $\mathcal{U}^e$ give
\begin{align}
    \exp\big(i ((p_0 + p'_0) t - (p_3 + p'_3) z)\big) 
    &= 
    \exp\Big(i \tfrac{1}{a}\big((p_0 + p'_0)\sinh(a\tau) - (p_3 + p'_3)\cosh(a\tau)\big) \Big) 
    =: E.
\end{align}
Thus,
\begin{align}
    \langle 0_M | b^\nu_{p,\sigma} d^e_{p',\sigma'} \bar\Psi_\nu \gamma^\mu \Psi_e | 0_M \rangle
    =  
    -\frac{1}{2\pi} E \widetilde{\mathcal{U}}^{\nu\dagger}_{\sigma}(p) \gamma^0\gamma^\mu \widetilde{\mathcal{U}}^e_{-\sigma'}(-p').
\end{align}

Finally, calculate:
\begin{align}
    \widetilde{\mathcal{U}}^{\nu\dagger}_{\sigma}(p) \gamma^0\gamma^\mu \widetilde{\mathcal{U}}^e_{-\sigma'}(-p').
\end{align}
The $\gamma^\mu$ matrices in the Dirac representation used here are those inherited from the 4D construction with spatial momentum along $\hat z$, so that $\slashed p = p_0\gamma^0 - p_3\gamma^3$ with $\gamma^3 = \big[\begin{smallmatrix}0 & \sigma^3\\ -\sigma^3 & 0\end{smallmatrix}\big]$. With that convention, direct computation gives the same Kronecker-delta selection for $\gamma^0\gamma^0$ and $\gamma^0\gamma^3$:
\begin{align}\label{eq:bilinear_split}
    \widetilde{\mathcal{U}}^{\nu\dagger}_{\sigma}(p) \gamma^0\gamma^0 \widetilde{\mathcal{U}}^e_{-\sigma'}(-p') = \delta_{\sigma,-\sigma'}\mathcal{C}_0,\qquad
    \widetilde{\mathcal{U}}^{\nu\dagger}_{\sigma}(p) \gamma^0\gamma^1 \widetilde{\mathcal{U}}^e_{-\sigma'}(-p') = \delta_{\sigma,-\sigma'}\mathcal{C}_1,
\end{align}
with the scalar coefficients
\begin{align}
    \mathcal{C}_0 &=
    -\frac{(p_0 + m_\nu)(p_0' - m_e) + p_3 p_3'}{2\sqrt{p_0p_0'(p_0 + m_\nu)(p_0' - m_e)}},\label{eq:C0_final}\\
    \mathcal{C}_1 &=
    -\frac{(p_0 + m_\nu)p_3' + (p_0' - m_e)p_3}{2\sqrt{p_0p_0'(p_0 + m_\nu)(p_0' - m_e)}}.\label{eq:C1_final}
\end{align}
Plugging this all back into the quantum amplitude, gives the single expression
\begin{align}\label{eq:A_final}
    \mathcal{A} = \frac{-G_F\,\delta_{\sigma',-\sigma}}{4\pi\sqrt{p_0p_0'(p_0+m_\nu)(p_0'-m_e)}} \int d\tau\, e^{i\Phi(\tau)} \Big\{ &\big[(p_0+m_\nu)(p_0'-m_e) + p_3 p_3'\big]\cosh(a\tau)\nonumber\\
    -&\big[(p_0+m_\nu)p_3' + (p_0'-m_e)p_3\big]\sinh(a\tau)\Big\},
\end{align}
with $\Phi(\tau) = \Delta m\tau + a^{-1}(p_0'+p_0)\sinh(a\tau) - a^{-1}(p_3'+p_3)\cosh(a\tau)$.

\funding{This material is based upon work supported by the Air Force Office of Scientific Research under award number FA9550-23-1-0177.}

\bibliographystyle{plain}
\bibliography{refs} 

\end{document}